\documentclass{PoS}
\newcommand\mx[1]{M_{\mathrm{#1}}}
\newcommand\mt{\mx{T}}
\newcommand\mzero{\mx{0}}
\newcommand\Nf{N_{\mathrm{f}}}

\title{Large mass hierarchies from strongly-coupled dynamics}

\ShortTitle{Large mass hierarchies from strongly-coupled dynamics}

\author{Andreas Athenodorou$^{ab}$, \speaker{Ed Bennett}\footnote{E-mail: {\tt e.j.bennett@swansea.ac.uk}} $^{cd}$, Georg Bergner$^e$, Daniel Elander$^f$, C.-J.\ David Lin$^{gh}$, Biagio Lucini$^c$, and Maurizio Piai$^c$\\
\llap{$^a$}Department of Physics, University of Cyprus, POB 20537, 1678 Nicosia, Cyprus\\
\llap{$^b$}Computation-based Science and Technology Research Center, The Cyprus Institute, 20 Kavafi Str., Nicosia 2121, Cyprus\\
\llap{$^c$}College of Science, Swansea University, Singleton Park, Swansea SA2 8PP, UK\\
\llap{$^d$}Kobayashi-Maskawa Institute for the Origin of Particles and the Universe (KMI), Nagoya University, Nagoya 464-8602, Japan\\
\llap{$^e$}Universit\"at Bern, Institut f\"ur Theoretische Physik, Sidlerstr. 5, CH-3012 Bern, Switzerland\\
\llap{$^f$}National Institute for Theoretical Physics, School of Physics, and Mandelstam Institute for Theoretical Physics, University of the Witwatersrand, Johannesburg, Wits 2050, South Africa\\
\llap{$^g$}Institute of Physics, National Chiao-Tung University, Hsinchu 30010, Taiwan\\
\llap{$^h$}CNRS, Aix Marseille Université, Université de Toulon, Centre de Physique Théorique, UMR 7332, F-13288 Marseille, France}

\abstract{Motivated by the absence of signals of new physics at the LHC, which seems to imply the presence of large mass hierarchies, we investigate the theoretical possibility that these could arise dynamically in new strongly-coupled gauge theories extending the standard model of particle physics. To this purpose, we study lattice data on non-Abelian gauge theories in the (near-)conformal regime---specifically, $\mathrm{SU}(2)$ with $N_{\mathrm{f}}=1$ and $2$ dynamical fermion flavours in the adjoint representation. We focus our attention on the ratio $R$ between the masses of the lightest spin-2 and spin-0 resonances, and draw comparisons with a simple toy model in the context of gauge/gravity dualities. For models in which large anomalous dimensions arise dynamically, we show indications that this mass ratio can be large, with $R > 5$. Moreover, our results suggest that $R$ might be related to universal properties of the IR fixed point. Our findings provide an interesting step towards understanding large mass ratios in the non-perturbative regime of quantum field theories with (near) IR conformal behaviour.}

\FullConference{34th annual International Symposium on Lattice Field Theory\\
		 24-30 July 2016\\
		 University of Southampton, UK}

\begin{document}

\section{Introduction}
The Higgs particle \cite{Aad:2012tfa,Chatrchyan:2012xdj} is the first example in nature of a boson the mass of which is not protected by symmetry arguments. Its low-energy Effective Field Theory (EFT) description in terms of a weakly-coupled scalar field is fine-tuned, as additive renormalisation makes it sensitive to unknown physics up to high scales. This is the hierarchy problem, one of the main motivations to investigate theoretical extensions of the Standard Model (SM).

Searches at the LHC currently show no new states not predicted by QCD \cite{Khachatryan:2016yec,ATLAS:2016eeo}, suggesting that should such states exist (and have non-vanishing couplings to the SM), then they must have masses large enough to have escape detection so far. If a new strongly-coupled theory is responsible for electroweak symmetry breaking and all the physical phenomena connected with it (reviewed in \cite{Piai:2010ma}), it would provide an elegant and conclusive solution to the hierarchy problem(s) of the electroweak theory; however, such a theory would need to exhibit a large \emph{mass hierarchy} between the would-be Higgs and other states of the theory.

We want to identify models that dynamically produce a large mass hierarchy between composite states, that cannot be explained in simple terms by symmetry arguments in a low energy EFT context (and without fine-tuning). We are aiming at something more than what in QCD is captured by the chiral Lagrangian, or Heavy Meson Chiral Perturbation Theory ($\chi$PT), that explain the masses and properties of pions and of heavy-light mesons, respectively. We want to find appropriate physical observables that allow such an identification to be assessed in a clean, unambiguous way distinctive from simple arguments formulated at weak coupling on the basis of internal symmetries.

In these proceedings, we provide one interesting step in this direction, and suggest possible ways to further develop this challenging research program in the future. Our starting point is identifying an observable that we can use to characterise a hierarchy in a strongly-coupled theory. Such a quantity would need to have a value near 1 for QCD, which does not exhibit such a hierarchy, and a significantly larger value for theories of phenomenological interest.

To this end we observe that, irrespectively of the microscopic details, all Lorentz-invariant four-dimensional field theories admit a stress-energy tensor $T_{\mu\nu}$. Correlation functions involving $T_{\mu\nu}$ can be analysed in terms of their scalar (trace) part and tensor (transverse and traceless) part. A well-defined observable is the ratio
\[
	R = \frac{\mt}{\mzero}
\]
where $\mt$ is the mass of the lightest spin-2 composite state, while $\mzero$ is the mass of the lightest spin-0 state. This quantity is defined universally, it can be computed explicitly in a wide variety of models, it is scheme-independent, and it is not directly controlled by internal global symmetries of the theory. It is legitimate to compare $R$ computed in theories with completely different internal symmetries and symmetry-breaking patterns. This is a particularly welcome feature in the context of gauge theories with fermionic field content, where the physics of chiral symmetry and its breaking introduces non-trivial model-dependent features.

\section{Mass deformation and infrared behaviour}
\begin{figure}
	\begin{center}
		\includegraphics[width=0.7\textwidth]{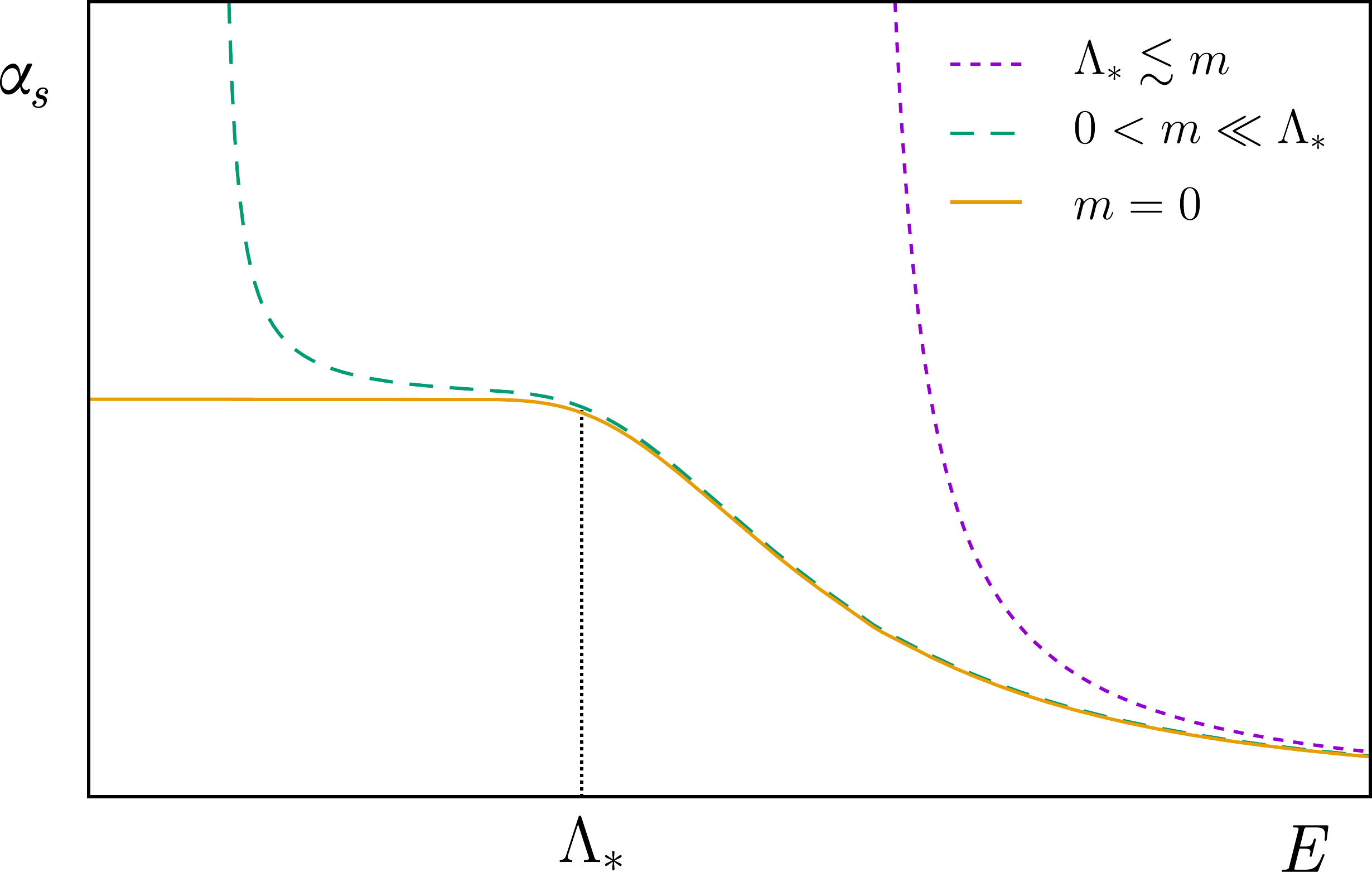}
	\end{center}
	
	\null\vspace{-24pt}\caption{Schematic representation of $\alpha_\mathrm{s} = g^2 / (4\pi)$ as a function of $E$ for three representative choices of fermion mass $m$. In yellow the massless $m = 0$ case, in which the only scale is $\Lambda_*$. In purple a case in which the mass $m$ is very large. In green a case in which $m \ll \Lambda_*$.\label{fig:deforming}}
\end{figure}

To better understand the analysis we will perform, we must have an idea of how $R$ will vary for a conformal theory as it is deformed by some mass $m$. In the absence of the deforming mass, the theory has a single scale $\Lambda_*$; the mass distorts the IR behaviour. The consequences of this are sketched in Fig.~\ref{fig:deforming}. For $m \gtrsim \Lambda_*$, the large mass limit (plotted in purple), the scale $\Lambda_*$ is inaccessible, so the theory is indistinguishable from one that confines. In the absence of a deforming mass, the theory is IR conformal, as shown in yellow. In the intermediary range meanwhile, the two scales are both visible---a crossover region is seen where the conformal behaviour is visible, before the confining behaviour takes over at scales near the deforming mass.

This allows us to probe how the spectrum will depend on $m$. In the region of small, finite deforming mass, then spectral masses will scale as 
	\[M \propto m^{1/\Delta}\;,\]
where $\Delta$ is the scaling dimension, equivalent to $1+\gamma^*$, where $\gamma^*$ is the anomalous dimension. Since the finite lattice volume introduces an absolute IR cutoff, then an appropriate scaling variable to consider is 
	\[x=Lm^{1/\Delta}\;.\] 
and so the $i$th state mass scales as
	\[LM_i = f_i(x)\;.\]
To first-order for a state of mass $M_0$, this gives
	\[L\mzero \propto x\;.\] 
which we can then substitute back in to give
	\[LM_i = f_i(L\mzero)\;.\]
Thus if we take mass ratios, then the dependence on the lattice extent drops out, allowing us to compare data at different volumes:
	\[\frac{M_i}{M_j} = \frac{f_i(L\mzero)}{f_j(L\mzero)}\]

We therefore expect four regimes for the ratio $R$. At large $m$, then the physics of interest is quenched out, so $R$ should become consistent with the value seen in pure Yang--Mills theory. For SU(2), this value has been found in lattice studies to be $R=1.44(4)$ \cite{Lucini:2004my}. Meanwhile at small $m$, the regime will depend on the lattice extent $L$. At sufficiently large volumes, then the observed physics includes the small $E$ region that shows confining properties. At very small volumes, then the scale $\Lambda_*$ is no longer observable, and so the theory observed is that of the ``femto-universe''. Previous studies indicate that $R \approx 1$ in this region \cite{GonzalezArroyo:1988dz,Daniel:1989kj,Daniel:1990iz}. However, there is an intermediary region of $L$ where it is sufficiently large to observe the conformal behaviour, without being so large as to be distorted by the confining effects. This is the region of interest, which can be extrapolated to a chiral limit value, which allows $R$ to be determined for the conformal case.

\section{Predictions from gauge--gravity duality}
Since we have constructed $R$ to allow comparison between disparate theories, we choose to explicitly draw this comparison. As a reference theory, we choose a toy model constructed making use of the principles of gauge--gravity duality. While full details of the model are given in \cite{Athenodorou:2016ndx}, the important fact to note here is that it is constructed within the bottom-up approach to holography, in such a way that the only operator deforming the theory and driving it away from conformality has scaling dimension $\Delta$. Thus for a theory investigated on the lattice found to have a particular scaling dimension, we may compare to the toy model with the corresponding value of $\Delta$.

\begin{figure}
	\begin{center}
		\includegraphics[width=0.7\textwidth]{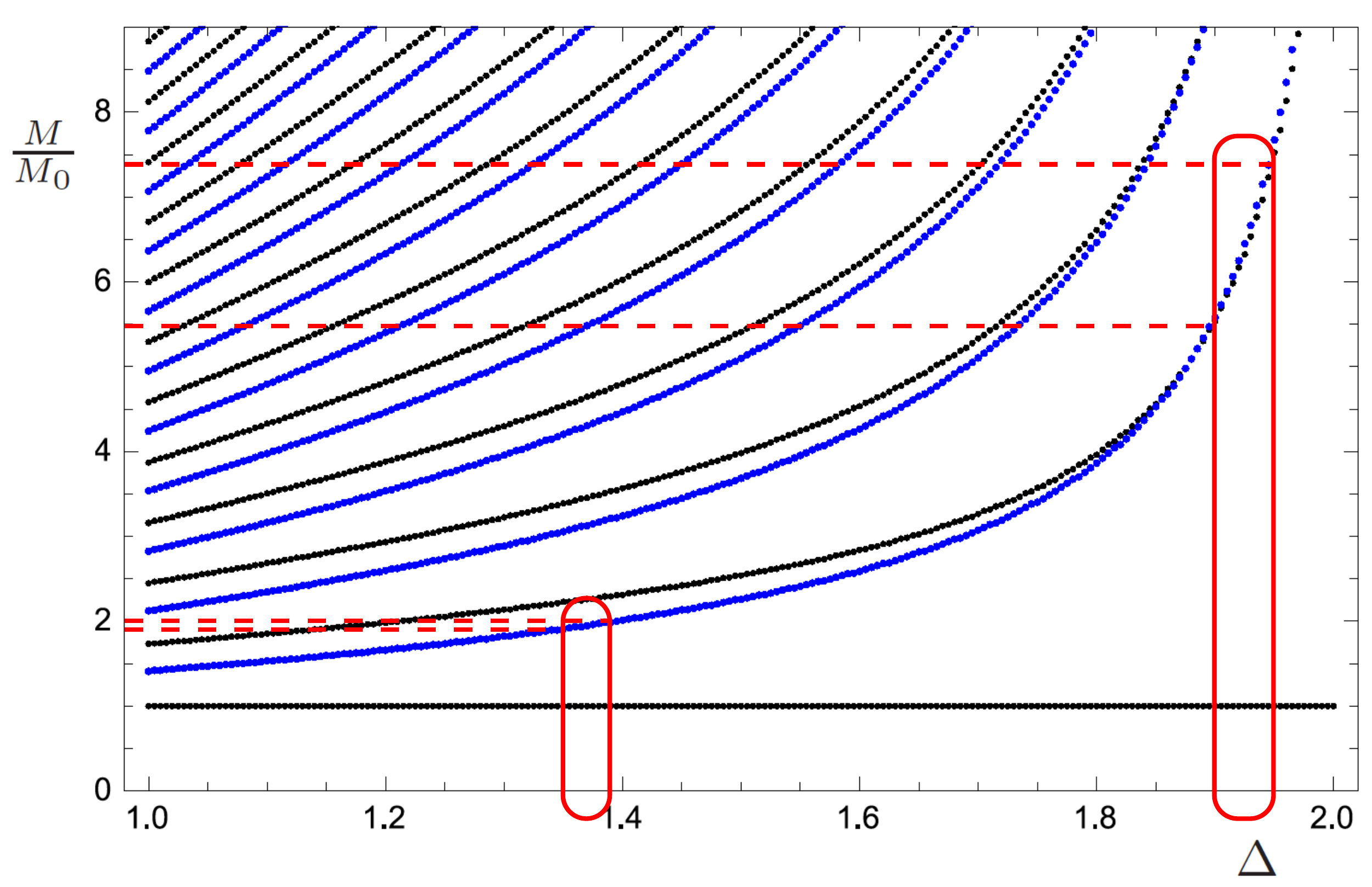}
	\end{center}
	\null\vspace{-36pt}\caption{The mass $M$ of the composite spin-0 (black) and spin-2 (blue) states, as a function of the scaling dimension $\Delta$. Masses are normalised by the mass of the lightest scalar $\mzero$, thus the lowest-lying blue line represents the ratio $R$. The relevant regions for the two theories considered are shown in red. \label{fig:gaugegravity}} 
\end{figure}

The methodology used allows for the full spectrum of composite spin-0 and spin-2 states to be computed numerically as a function of $\Delta$. By taking the ratio of the lowest-lying state for each case, we may then find the value of $R$ predicted from the gauge--gravity theory. The main result of the gravity analysis is that the value of $R$ is a monotonically increasing function of $\Delta$, as shown by the spectrum of scalar and tensor states in Fig.~\ref{fig:gaugegravity}.

\section{Lattice results}
We consider lattice results from two theories: SU(2) gauge theory with one and two flavours of adjoint Dirac fermions. Both have been studied on the lattice using the Wilson plaquette action and the Wilson fermion action, and found to show signals of being in or near the conformal window \cite{Athenodorou:2014eua,DelDebbio:2010hx}. The two-flavour theory has been found (at $\beta=2.25$) to have anomalous dimension $\gamma^* = 0.371(20)$ \cite{DelDebbio:2015byq}, corresponding to $\Delta=1.371$. The one-flavour theory has been found (at $\beta=2.05$) to have anomalous dimension $\gamma^* = 0.925(25)$, corresponding to $\Delta=1.925(25)$ \cite{Athenodorou:2014eua}. The relevant region of the gauge--gravity plot for each range of $\Delta$ is shown in Fig.~\ref{fig:gaugegravity} as the left and right red highlight respectively, giving values $R_{\Nf=2}\approx 1.95(4)$ and $R_{\Nf=1} \approx 6.53^{+1.50}_{-0.91}$.

\begin{figure}
	\begin{center}
		\includegraphics[width=0.7\textwidth]{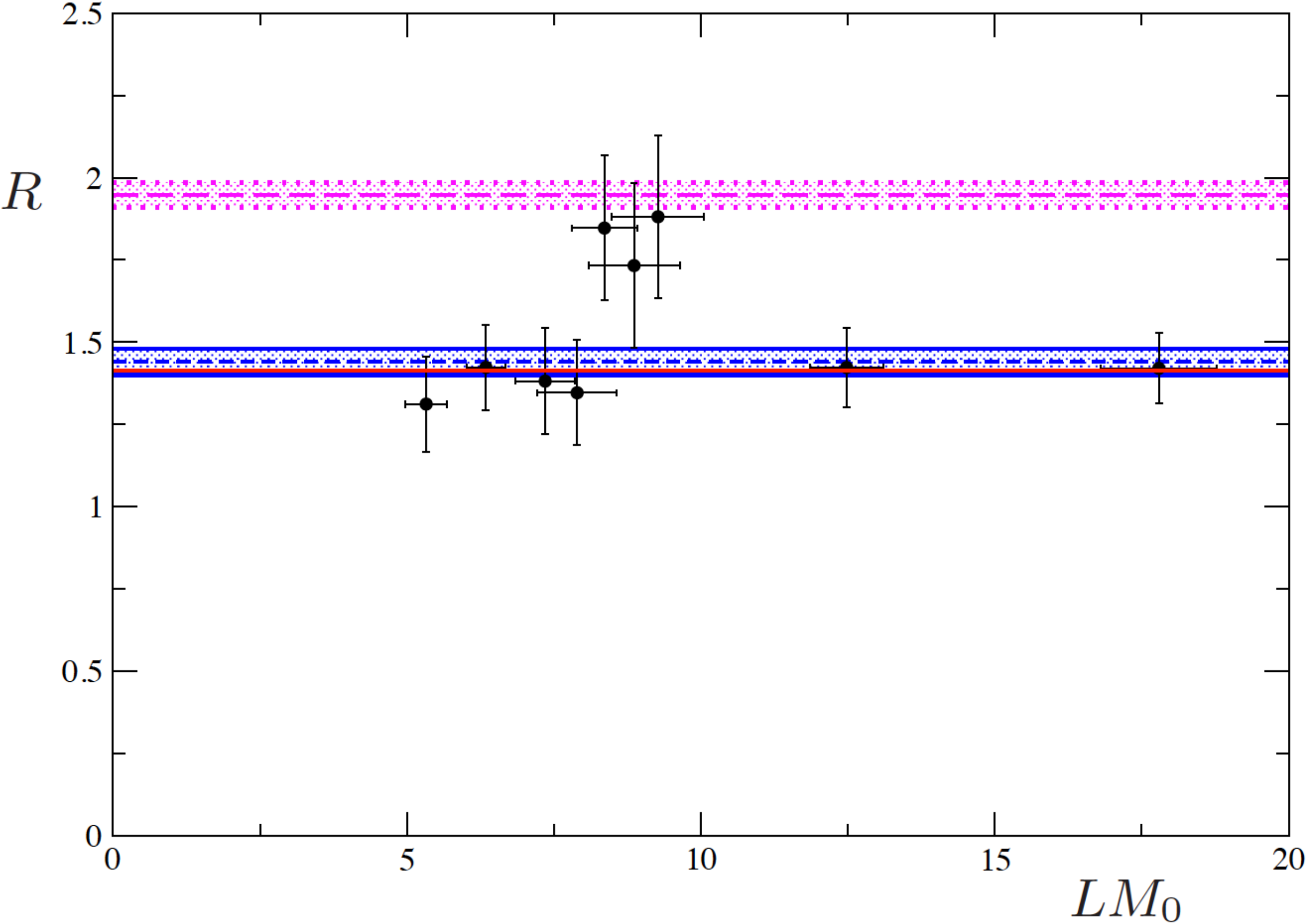}
	\end{center}
	
	\null\vspace{-24pt}\caption{The ratio $R = \mt / \mzero$ computed for SU(2) with two adjoint Dirac flavours (black points). The pink and blue bands, and the red line, are described in the text.\label{fig:nf2}}
\end{figure}
\begin{figure}
	\begin{center}
		\includegraphics[width=0.7\textwidth]{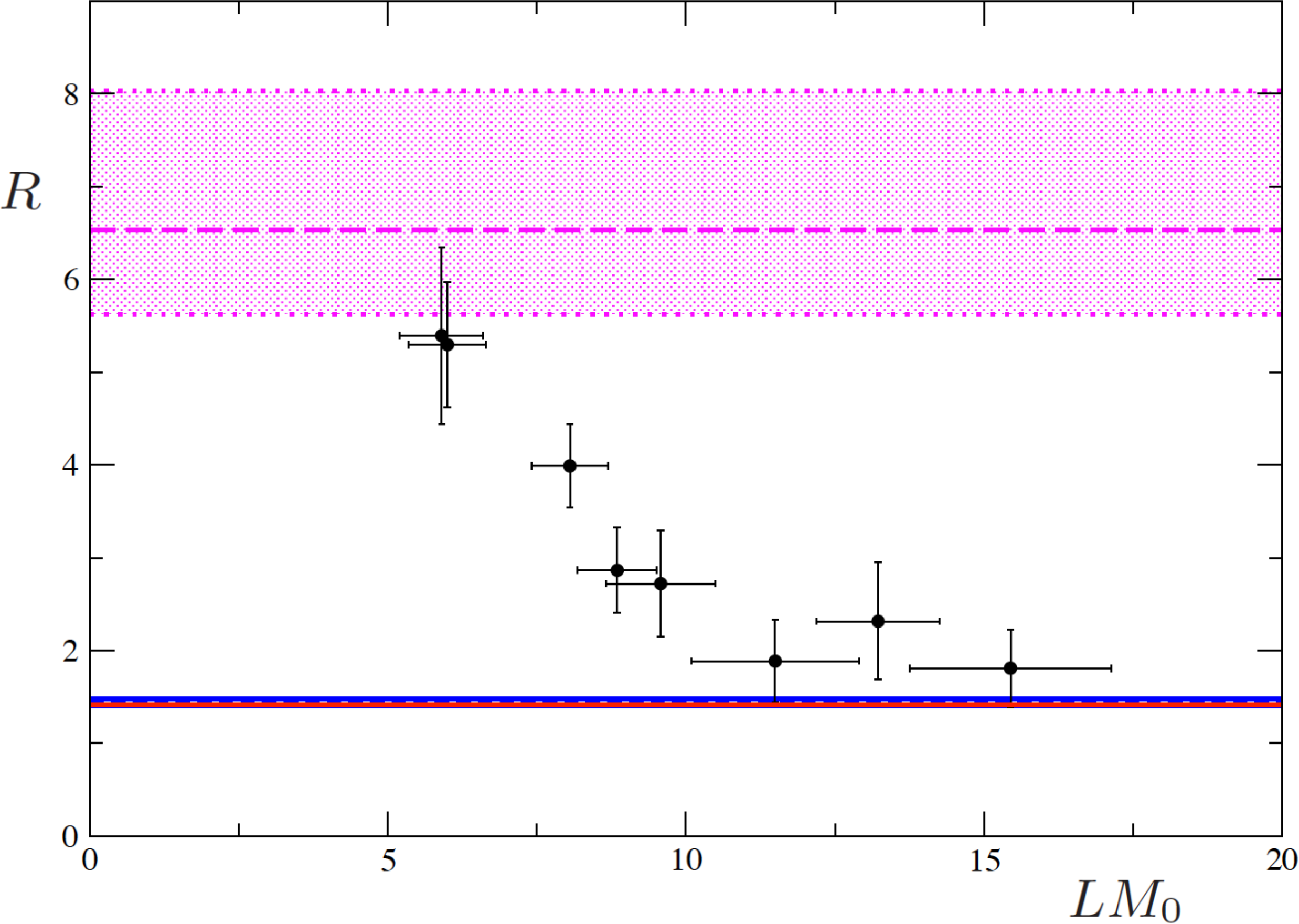}
	\end{center}
	
	\null\vspace{-24pt}\caption{The ratio $R = \mt / \mzero$ computed for SU(2) with one adjoint Dirac flavour (black points). The pink and blue bands, and the red line, are described in the text.\label{fig:nf1}}
\end{figure}

In each case we make use of configurations from the existing lattice studies mentioned above, which were not optimised for this study. As such a range of both deforming mass $m$ and lattice extent $L$ were considered. The results for $R$ found in both cases are plotted in Figs.~\ref{fig:nf2} and \ref{fig:nf1}. In each case, the prediction from the gauge--gravity model at corresponding $\Delta$ is shown as a pink band, while the numerical result for pure-gauge SU(2) is shown as a blue band. The red line is the result from the GPPZ model \cite{Girardello:1999bd}, on which the gauge--gravity model is based, which has $\Delta=1$ and $R=\sqrt{2}$.

In the two-flavour case, at large values of $L\mzero$, $R$ closely matches the pure gauge value---as anticipated at large $\mzero$. At small values of $L\mzero$, $R$ is again near the pure-gauge value but shows signs of descending below it as predicted by the femto-universe; this is expected for small values of $L$. In the intermediary range, $R$ approaches (and takes values compatible with) the predictions of the gauge--gravity model for a theory with the same scaling dimension.

Meanwhile in the one-flavour case, at large values of $L\mzero$, the behaviour is similar to the two-flavour theory---$R$ is compatible with the pure gauge result. Once again, as $L\mzero$ gets smaller, the ratio $R$ gets larger, and once again enters a range compatible with the predictions of the gauge--gravity model at the same scaling dimension. However in this case we do not see the other side of this peak, heading towards the femto-universe.

\section{Conclusions}
We have calculated the measure $R=\mt/\mzero$ for a variety of theories, both lattice gauge theories with adjoint matter and toy gauge--gravity models. We see an unexpected degree of agreement between the two when the scaling dimension is held constant; if this were a hint of universality, then it would suggest that theories of phenomenological interest that have a large anomalous dimension $\approx 1$ would also by necessity exhibit the large hierarchy of scales necessary to explain their absence in LHC searches.

A dedicated study of each theory is necessary, performing scans separately of $L$ and $m$, to ensure that the observed behaviour is not an artefact. Further work is also needed to probe other theories of phenomenological interest and investigate whether the observed hints of possible universality are also observed there.

\section*{Acknowledgements}
Numerical computations were executed in part on the HPC Wales systems, supported by the ERDF through the WEFO (part of the Welsh Government), on the Blue Gene Q system at the Hartree Centre (supported by STFC) and on the DiRAC Blue Gene/Q Shared Petaflop system at the University of Edinburgh. The latter is operated by the Edinburgh Parallel Computing Centre on behalf of the STFC DiRAC HPC Facility (www.dirac.ac.uk). This equipment was funded by BIS National E-infrastructure capital grant ST/K000411/1, STFC capital grant ST/H008845/1, and STFC DiRAC Operations grants ST/K005804/1 and ST/K005790/1. DiRAC is part of the National E-Infrastructure.

\bibliographystyle{JHEP}
\bibliography{references.bib}

\end{document}